\documentclass[iop]{emulateapj}
\usepackage{epsfig}
\usepackage{subfigure}
\usepackage{mathtools}
\usepackage{bm}
\usepackage{amsmath}
\usepackage{hyperref}

\def\Kepler{\textit{Kepler}}

\def\vsini{V\sin I_\star}
\def\k{\kappa}
\def\L{\mathcal L}
\def\veq{V}
\def\prot{P_{\rm rot}}
\def\ltsima{$\; \buildrel < \over \sim \;$}
\def\lsim{\lower.5ex\hbox{\ltsima}}
\def\gtsima{$\; \buildrel > \over \sim \;$}
\def\gsim{\lower.5ex\hbox{\gtsima}}

\def\Ntot{70}
\def\Nsingle{45}
\def\Nmulti{25}

\begin{document}

\title{
Obliquities of Kepler stars: comparison of single- and multiple-transit systems
}

\author{Timothy D.\ Morton\altaffilmark{1} \& Joshua N.\ Winn\altaffilmark{2,3}}

\email{tdm@astro.princeton.edu}

\altaffiltext{1}{Department of Astrophysical Sciences, 4 Ivy Lane, Peyton Hall, Princeton University, Princeton, NJ 08544}

\altaffiltext{2}{Physics Department, Massachusetts Institute of Technology, 77 Mass.\ Ave., Cambridge, MA 02139}

\altaffiltext{3}{MIT Kavli Institute for Astrophysics \& Space Research, 70 Vassar St., Cambridge,
MA 02139}

\begin{abstract}

  The stellar obliquity of a transiting planetary system can be constrained by combining measurements of the star's rotation period, radius, and projected rotational velocity. Here we present a hierarchical Bayesian technique for recovering the obliquity distribution of a population of transiting planetary systems, and apply it to a sample of 70 \textit{Kepler} Objects of Interest.
With $\approx$95\% confidence
we find that the obliquities of stars with only a single detected transiting planet are systematically larger than those with multiple detected transiting planets. This suggests that a substantial fraction of \textit{Kepler}'s single-transiting systems represent dynamically hotter, less orderly systems than the ``pancake-flat'' multiple-transiting systems.

\end{abstract}

\section{Introduction}

At least half of Sun-like stars have a planet with a period shorter than Mercury's 88-day period \citep{mayor2011,fressin2013}, and in many cases there is more than one such planet. It would be interesting to know whether these compact multiplanet systems formed in a fundamentally different way from that of other types of planetary systems, such as hot Jupiters or the Solar System. Clues can be obtained by comparing the systems' geometrical parameters. For example, there is growing evidence that compact multiplanet systems generally have coplanar orbits, similar to the Solar System \citep{fang2012,figueira2012,johansen2012,tremainedong2012,fabrycky2014}.

Another geometric parameter is the stellar obliquity, the angle between the angular momentum vectors of the host star's rotation and the orbit of one of its planets. The Sun's obliquity is $7^\circ$ relative to the ecliptic. Obliquities have been measured for dozens of exoplanet host stars, and have been found to range widely from smaller than a few degrees to nearly 180$^\circ$ [see, e.g., \citet{winn10, triaud10, albrecht2012}, or the online compilation by R.\ Heller\footnote{\url{http://www.physics.mcmaster.ca/~rheller/}}]. Most of these measurements have been for hot Jupiters and were based on the Rossiter-McLaughlin effect, the distortion of stellar absorption lines that appears during a planetary transit. Because this technique relies on precise spectroscopy of transits, it is harder to apply to the known population of compact multiplanet systems, which tend to involve fainter stars, smaller transit depths, and less frequent transits. Other techniques are being developed, such as the analysis of starspot-crossing anomalies and asteroseismology, which have enabled a few obliquity measurements for multiplanet systems \citep{sanchisojeda2012, hirano2012a, albrecht2013, chaplin2013, huber2013,vaneylen2014}.

This paper presents a statistical technique for comparing the stellar obliquity distributions of different samples of exoplanetary systems, based on measurements of the stars' sky-projected rotation velocities ($\vsini$). The idea is that transit-hosting stars with anomalously low values of $\vsini$ are likely to have high obliquities, because the star is likely to have low $\sin I_\star$ whereas the orbits of transiting planets necessarily have high $\sin I_P$. The advantage of this technique is that it less observationally demanding, requiring only a single and non-time-critical spectrum to obtain $\vsini$. The disadvantage is that it provides relatively coarse statistical information rather than precise individual measurements.

This method was put into practice by \cite{schlaufman2010}, who applied it to hot Jupiter systems, and by Hirano et al.\ (2012b, 2014) and \cite{walkowicz2013}, who applied it to {\it Kepler} systems.  {\it Kepler} systems have the advantage that the stellar rotation period $P_{\rm rot}$ can sometimes be measured from quasiperiodic flux variations, allowing the $V$ in $\vsini$ to be estimated directly as $2\pi R_\star/P_{\rm rot}$ rather than using gyrochronology or other indirect means to predict $V$. In particular, \citet{hirano2014} performed the most sophisticated analysis of {\it Kepler} systems to date. They calculated posterior probability distributions for $I_\star$ on a star-by-star basis, and used the results to demonstrate that their sample of 25 stars is inconsistent with an isotropic obliquity distribution; there is a tendency toward spin-orbit alignment. However, based on several other statistical tests, they did not find strong evidence for any difference in the obliquity distributions of stars with single and multiple transiting planets.

Here we describe a potentially more powerful statistical framework for comparing obliquity distributions of different samples of exoplanetary systems (\S~2), test it on simulated data (\S~3) and apply it to a larger sample than was considered previously (\S~4). We find evidence that stars with multiple transiting planets have systematically lower obliquities than stars with only a single transiting planet, which further suggests that the compact multiple-transiting systems are a separate group that formed through a different mechanism (\S~5).

\section{Formalism}

We assume the planet's orbit has $\sin I_P \approx 1$, and denote the stellar obliquity by $\theta$. Following \citet{fabrycky2009} we model the obliquity distribution as a Fisher
distribution\footnote{Eqn.~(\ref{eq:fisher} is a special case ($p=3$) of the more
general von Mises-Fisher distribution, which was studied in detail by \citet{fisher1953}.},
\begin{equation}
\label{eq:fisher}
f_{\theta} \left(\theta | \k \right) = \frac{\k}{2 \sinh \k} \exp \left(\k \cos \theta \right) \sin \theta.
\end{equation}
\cite{tremainedong2012} also used this distribution---analogous to a zero-mean normal distribution on a sphere---to model the mutual inclination distribution of exoplanetary systems.  The parameter $\k$ gives the degree of concentration of the distribution; for large $\k$, the distribution becomes a Rayleigh distribution with width $\sigma = \k^{-1/2}$, and as $\k \rightarrow 0$, the distribution becomes isotropic.

We consider $N$ stars for which $\vsini$, $P_{\rm rot}$ and $R_\star$ have been measured, resulting in $N$ posterior probability distributions $\{I_\star\}$.  
The knowledge of each posterior probability distribution may take the form of $K$ samples. 
We wish to calculate the posterior probability distribution for $\k$. For reasons that will become clear, we express this in terms of $\cos I_\star$ rather than $I_\star$ itself:
\begin{equation}
\label{eq:bayeskappa}
p_\k \left(\k | \{\cos I_\star\} \right) \propto \L_\k \left( \{\cos I_\star\} | \k \right) \pi_\k (\k),
\end{equation}
where $\L_\k$ is the likelihood function for $\k$ conditioned on the data, and $\pi_\k$ is the prior.  We adopt the same uninformative prior that was proposed and justified by \citet{fabrycky2009}:
\begin{equation}
\label{eq:kappaprior}
\pi_\k(\k) \propto \left(1 + \k^2\right)^{-3/4}.
\end{equation}

Our framework for testing for differences between obliquity distributions is based on that of \citet{hogg2010}, who gave a hierarchical Bayesian prescription for inferring the parameters of the underlying population distribution of a desired quantity, given a set of posterior samplings of that quantity. An excellent and succinct
pedagogical description of this technique was given recently by
\cite{dfm2014}. Based on Eqn.~(9) of \citet{hogg2010}, the likelihood function for $\k$ may be approximated as
\begin{equation}
\label{eq:hogg}
\L_{\k} \approx \prod_{n=1}^{N} \frac{1}{K} \sum_{k=1}^K \frac{f_\k(c_{nk})}{\pi_0(c_{nk})}.
\end{equation}
Here, the product is over the $N$ stellar inclination measurements, and the sums inside the product are over $K$ posterior samplings. To cast this equation in a form similar
to that of \citet{hogg2010}, we have used $c_{nk}$ as an abbreviation
for the $k$th sample of the $n$th posterior for $\cos I_\star$.
Likewise, $f_\k$ is an abbreviation
for $f_{\cos I_\star}(\cos I_\star | \kappa)$, the probability density for $\cos I_\star$ given a value of $\k$.
Finally, the function $\pi_0$ is the original uninformative prior on $\cos I_\star$ upon which the posterior samplings were based. Therefore, to calculate this likelihood, the required ingredients are
(i) the function representing the prior on $\cos I_\star$,
(ii) the probability distribution function for $\cos I_\star$ given $\k$,
and (iii) posterior samplings for the cosines of each of $N$ different stellar inclinations. We now discuss these in turn.

For the prior on $\cos I_\star$ we make the simple choice of a uniform distribution from 0 to 1, corresponding to an isotropic distribution on a sphere. This is the $\pi_0$ function in the denominator of Equation~(\ref{eq:hogg}).

\begin{figure}[t!]
   \centering
   \includegraphics[width=3.5in]{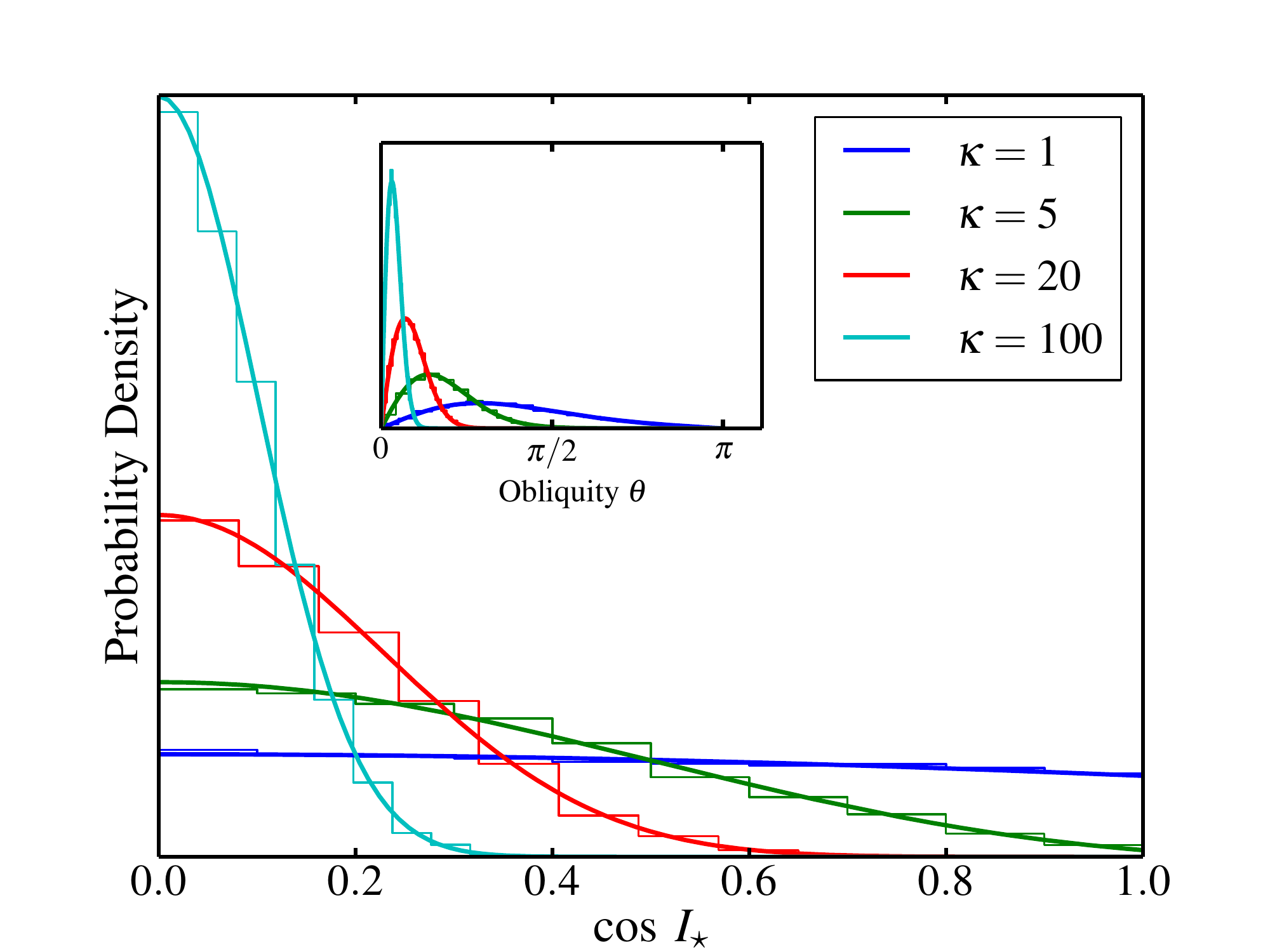} 
   \caption{Comparison of Monte Carlo simulations (histograms) to the analytic probability density of $\cos I$ derived here (Equation \ref{eq:fcosi}; solid curves), given different values of the Fisher parameter $\k$.  The obliquity $\theta$ is drawn from
a Fisher distribution (inset plot). }
      \label{fig:cosidistributions}
\end{figure}

To derive $f_\k$, the probability distribution for $\cos I_\star$ given $\k$, the first step is to relate the obliquity $\theta$ to the cosine of the line-of-sight stellar rotation inclination angle $I_\star$:
\begin{equation}
\label{eq:cosi}
\cos I_\star = \sin \theta \cos \phi,
\end{equation}
where $\phi$ is the azimuthal angle of the stellar rotation axis, using a polar coordinate
system for which the planet's orbital axis is the $z$-axis and $\phi=0$ along the line of sight.
We recognize the form of Eqn.~(\ref{eq:cosi}) as
\begin{equation}
\label{eq:ZeqXY}
Z = X Y,
\end{equation}
with $Z=\cos I_\star$, $X=\sin\theta$ and $Y=\cos\phi$.
If the probability distributions for $X$ and $Y$ are known to be $f_X$ and $f_Y$, then the probability distribution for $Z$ is \citep{rohatgi1976}
\begin{equation}
\label{eq:fZ}
f_Z(z) = \int_{-\infty}^{\infty} f_X(x) f_Y(z/x) \frac{1}{|x|} dx
\end{equation}
In our case we do not begin with expressions for $f_X$ or $f_Y$, but rather
with the fact that $f_\theta$ is a Fisher distribution (Eqn.~\ref{eq:fisher})
and $f_\phi = 1/(2\pi)$.
To obtain the distributions for $f_{\sin\theta}$ and $f_{\cos\phi}$, we use the following equation for the distribution of $Y = g(X)$ given that $f_X$ is known:\footnote{See, e.g., \url{http://en.wikipedia.org/wiki/Probability_density_function}}
\begin{equation}
\label{eq:fY}
f_Y(y) = \sum_{k=1}^{n(y)} \left| \frac{d}{dy} g^{-1}_{k}(y) \right| \cdot f_X[g^{-1}_{k}(y)],
\end{equation}
where the indexing is over the different solutions to $y = g(x)$ [which in this
case are the two solutions for $g(x) = \sin x$].
We are now in the position to write down the probability distributions for $\sin \theta$ and $\cos \phi$.  The distribution of the sine of an angle that is Fisher-distributed turns out to be 
\begin{equation}
\label{eq:fsintheta}
f_{\sin \theta}(y | \kappa) = \frac{\kappa}{\sinh{\kappa}} \frac{y}{\sqrt{1-y^2}} \cosh\left( \kappa \sqrt{1-y^2}\right).
\end{equation}
The distribution of cosine of the uniformly distributed angle $\phi$ is
\begin{equation}
\label{eq:fcosphi}
f_{\cos \phi}(x) = \frac{2}{\pi \sqrt{1-x^2}}.
\end{equation}
For simplicity, both of these distributions are normalized to be valid on the interval [0,1), rather than ($-1$,1).

Combining these ingredients, we have an expression for the probability distribution for $\cos I_\star$ given a value of $\k$:
\begin{equation}
\label{eq:fcosi}
f_{\cos I_\star}(z | \kappa) = \frac{2 \kappa}{\pi \sinh \kappa}\int_z^1 \frac{\cosh\left( \kappa \sqrt{1-y^2} \right)}{\sqrt{1-y^2}} \frac{1}{\sqrt{1-\left(z/y\right)^2}}  dy.
\end{equation}
This is the function we abbreviated as $f_\k$ in Equation~(\ref{eq:hogg}).
The integral ranges from $z$ to 1 rather than $-\infty$ to $\infty$ because both $X$ and $Y$ are defined only from 0 to 1. This integral does not have an analytic solution and must be evaluated numerically. We confirmed the accuracy of this equation through direct Monte Carlo simulations, as shown in Figure~\ref{fig:cosidistributions}.

\begin{figure}[t!]
   \centering
   \includegraphics[width=3.5in]{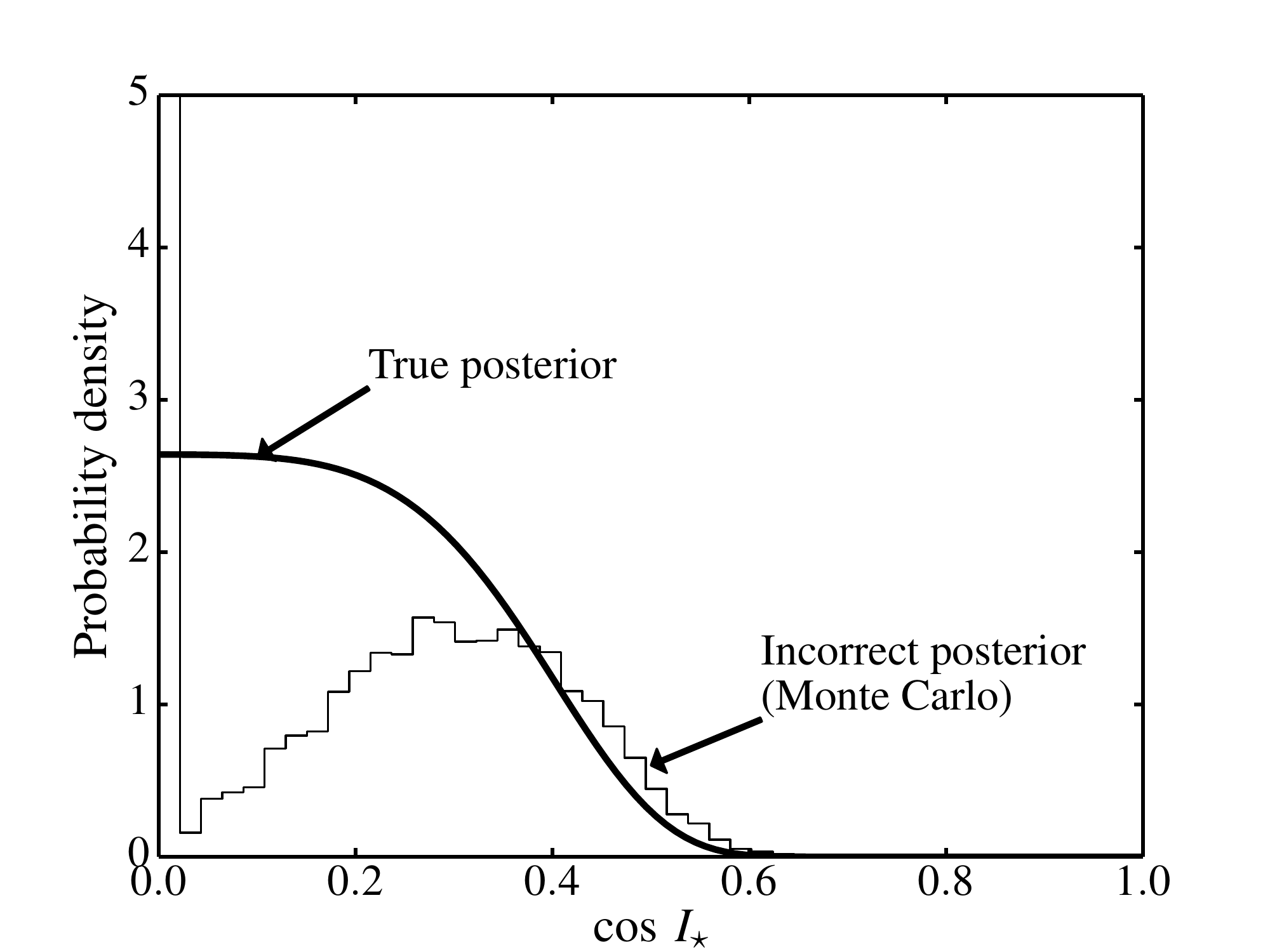} 
   \caption{
The posterior probability distribution for $\cos I_\star$ for a star with
$\vsini = V = 10 \pm 0.5$~km~s$^{-1}$, using Equations \ref{eq:cosiposterior} and \ref{eq:cosilhood} (thick curve). The histogram shows the faulty results of
the intuitively appealing but incorrect procedure
of generating posterior samples for $\sin I_\star$ by dividing samples from the $\vsini$ and $\veq$ posteriors and then converting to $\cos I_\star$. In the histogram, the pile-up
at $\cos I_\star=0$ (containing $\sim$50\% of the samples) corresponds to the cases when the ratio of $\vsini$ and $V$ exceeded unity
and was clipped to unity.
}
      \label{fig:cosiposterior}
\end{figure}

Now we can explain why we chose to express this problem in terms of $\cos I_\star$ rather than $I_\star$ or $\sin I_\star$. Eqn.~(\ref{eq:fcosi}) might appear complex, but it is actually simple in comparison to the equations that are obtained for the probability distributions of $I_\star = \cos^{-1} (\sin \theta \cos \phi)$ or $\sin I_\star = \sqrt{1 - (\sin \theta \cos \phi)^2}$ in terms of $\k$.

Finally, for each star we need samples from the posterior probability distributions for $\cos I_\star$, at which the $f_\k$ and $\pi_0$ functions are evaluated.  The simplest way to obtain these samples is to derive the posterior probability distribution for $\cos I_\star$ for each star, and then sample from those distributions.
We suppose the data $D$ takes the form of probability distributions for $\vsini$ and
equatorial rotational velocity $V = 2\pi R_\star/P_{\rm rot}$.
Then the posterior probability distribution for $\cos I_\star$ can be computed using
\begin{equation}
\label{eq:cosiposterior}
p(\cos I_\star | D) \propto \L \left(D | \cos I_\star \right).
\end{equation}

\begin{figure*}[t!]
\hfill
\subfigure[]{\includegraphics[width=3.5in]{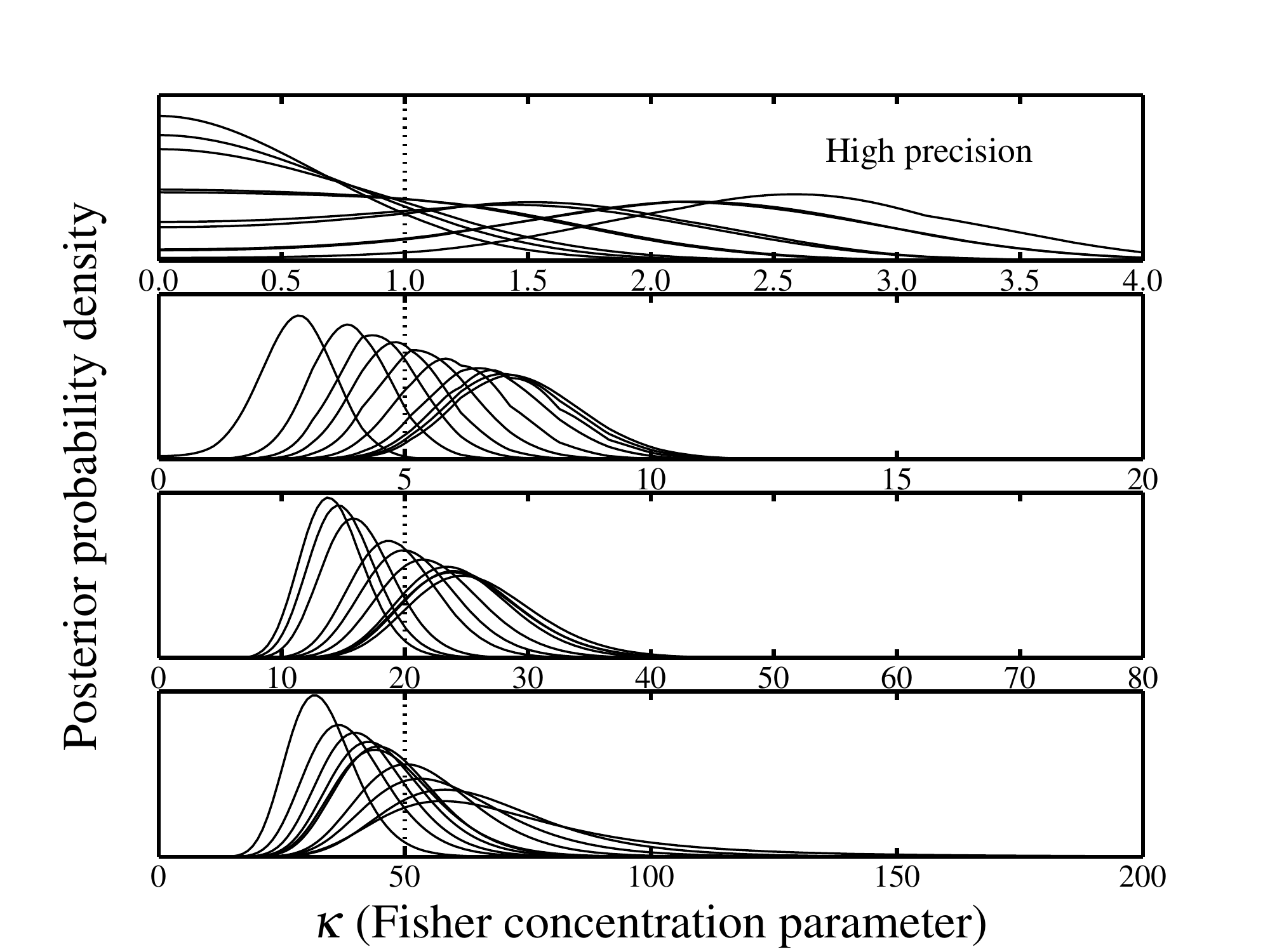}}
\hfill
\subfigure[]{\includegraphics[width=3.5in]{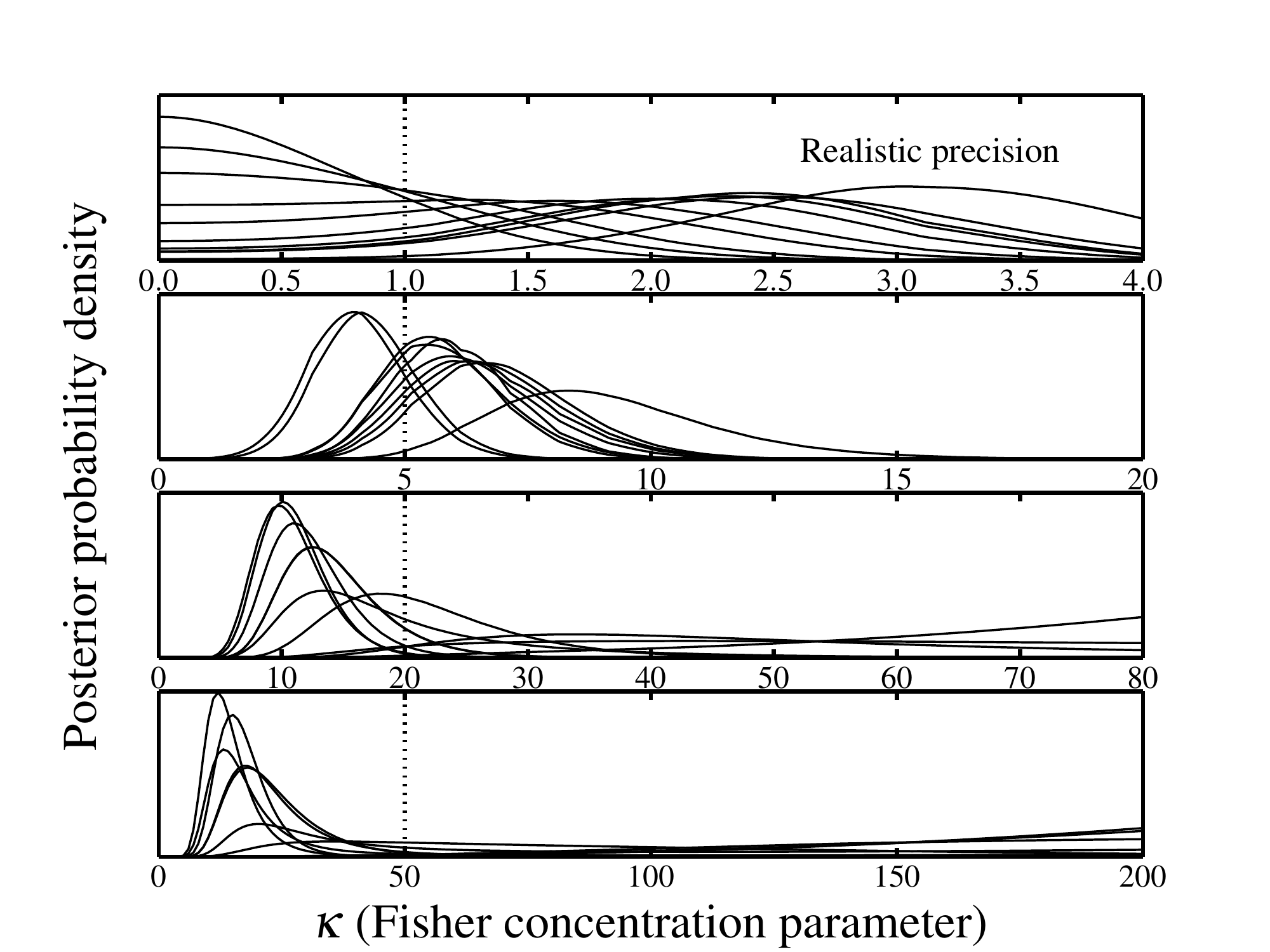}}
\hfill
\caption{Application to simulated data.
In all cases \Ntot\ stars are simulated, matching the number of actual
stars analyzed in \S~\ref{sec:real-data}.
In each frame, the dotted vertical line shows the true value of $\k$,
and the curves are the $\k$ posteriors resulting from the analysis
of 10 different Monte Carlo realizations of the simulated data.
{\it Left.}---High precision measurements are assumed, with
1$\sigma$ uncertainties of 1\% in $R_\star$ and $P_{\rm rot}$,
and 0.1~km~s$^{-1}$ in $\vsini$.
{\it Right.}---More realistic uncertainties are assumed:
10\% in $R_\star$, 3\% in $P_{\rm rot}$, and 0.5~km~s$^{-1}$ in $\vsini$.
Low values of $\k$ are recovered well.
In the realistic-uncertainty simulations, values of $\k \gtrsim 20$ (distribution concentrated toward low obliquity) can sometimes be mistaken for smaller values (a consequence of the prior on $\k$ we are using), but the reverse is not true.
}
\label{fig:simtests}
\end{figure*}

We have omitted the prior for $\cos I_\star$ on the right-hand side of this equation, because it is taken to be a constant. If the probability distribution for $\vsini$ is $p_{Vs}$ and the probability distribution for $V$ is $p_{V}$, we may write the likelihood function as follows:
\begin{equation}
\label{eq:cosilhood}
\L \left(D | \cos I_\star \right) = \int_0^\infty p_{Vs}(v) p_{V}\left( \frac{v}{\sqrt{1-\cos^2 I_\star}} \right) dv
\end{equation}   
For each star, this posterior can be constructed and then sampled to provide the samples at which to evaluate Eqn.~(\ref{eq:hogg}). The thick curve in Figure~\ref{fig:cosiposterior} shows an example of the posterior for $\cos I_\star$, for a case in which both $\vsini$ and $V$ are $10 \pm 0.5$~km~s$^{-1}$ with Gaussian distributions. The
posterior is flat for $\cos I_\star \lsim 0.2$ and declines sharply
for values larger than 0.4. The results are consistent with perfect alignment
($\cos I_\star = 0$, $\sin I_\star = 1$) as expected.

This figure also illustrates why Bayesian posterior estimation is important in this case. One might imagine obtaining samples from the $\cos I_\star$ posterior by dividing a set of Monte Carlo samples from the $\vsini$ posterior by a set of samples from the $V$ posterior to get a sampling of $\sin I_\star$.  The question would arise on what to do when this division gives a result exceeding unity; it would seem intuitive to simply set such samples equal to unity. Then, the samples of $\sin I_\star$ could be converted into samples of $\cos I_\star$.  This was the procedure used by \citet{hirano2014} and perhaps others in the past.  However, we find that the results of this intuitively appealing procedure match the true posterior poorly. The histogram in Figure \ref{fig:cosiposterior} shows the results of this incorrect procedure for our example with $\vsini = V = 10 \pm 0.5$~km~s$^{-1}$, which falsely suggest that the posterior for $\cos I_\star$ has a peak near 0.3 and a delta-function at $\cos I_\star = 0$.

\section{Application to simulated data}
\label{sec:simulated-data}

To test and demonstrate this formalism we apply it to a variety of simulated measurements of stellar inclination angles, using the following procedure:
\begin{enumerate}

\item Select $N$ stars randomly from the SPOCS catalog \citep{valenti2005}, which gives values of $T_{\rm eff}$, $\log g$, and $R_\star$, among other properties. We imagine each star has a transiting planet with $\sin I_\star=1$.

\item Assign a rotation period to each star appropriate for its effective temperature and surface gravity, based to the rotational evolution models of \citet{vansaders2013}.

\item Determining each star's rotation velocity, $V = 2\pi R_\star/P_{\rm rot}$.

\item Draw an obliquity $\theta$ for each star from a Fisher distribution with a given $\k$; draw an azimuthal angle $\phi$ from a uniform distribution from 0 to $2\pi$; and calculate the stellar inclination $I_\star = \cos^{-1} \left( \sin \theta \cos \phi \right)$.

\item Calculate $\vsini$ based on the previously determined values of $V$ and $\sin I_\star$.

\item Simulate measurements of each system: $\vsini$, $R_\star$, and $P_{\rm rot}$ are all assumed to have Gaussian uncertainty distributions. We test different choices for the 1$\sigma$ uncertainties in each parameter.

\item Calculate the posterior for $\cos I_\star$ for each star, using Eqn.~(\ref{eq:cosilhood}).

\item Sample these posteriors 1000 times each, and derive the posterior distribution for $\k$ using Equations~(\ref{eq:bayeskappa}) and (\ref{eq:hogg}).

\end{enumerate}
Figure \ref{fig:simtests} shows how well we can recover the true value of $\k$ from a sample of $N=70$~stars, depending on the precision with which $\vsini$, $R_\star$, and $P_{\rm rot}$ are measured. As expected, for very high precision (left panel) the true value is recovered well, and for more realistic assumptions about the measurement precision (right panel) the posteriors for $\k$ are broader.  

For the $\k=20$ and $\k=50$ tests with realistic uncertainties, the derived posteriors seem to give systematically low results for $\k$ compared to the input values.  We believe the primary reason for this behavior is the choice of prior for $\k$, given by Eqn.~(\ref{eq:kappaprior}). This function is peaked at small values, so as not to give undue prior weight to models with well-aligned distributions. To support this statement we repeated our Monte Carlo calculations using a uniform prior in $\k$, and did not find the same type of bias.  Specifically we found that for $\k=20$, six of the ten simulations have maximum-posterior values of $\k < 20$, and four have $\k > 20$; and for the $\k=50$ simulations, five have maximum-posterior values $<$50 and five $>$50.
(We also note that the apparent systematic bias of the posteriors seems to be exaggerated by an
optical illusion: the posteriors with significant weight at larger $\k$ are more spread out and have lower amplitudes, whereas those centered at smaller $\k$ values appear much more prominent.)

The conclusions of these test simulations are (a) our inference apparatus is working properly, and (b) given the prior we are using, values of $\k \gtrsim 20$ can sometimes be mistaken for smaller $\k$, whereas the reverse is not true.

\begin{figure*}[t!]
   \centering
   \includegraphics[width=6in]{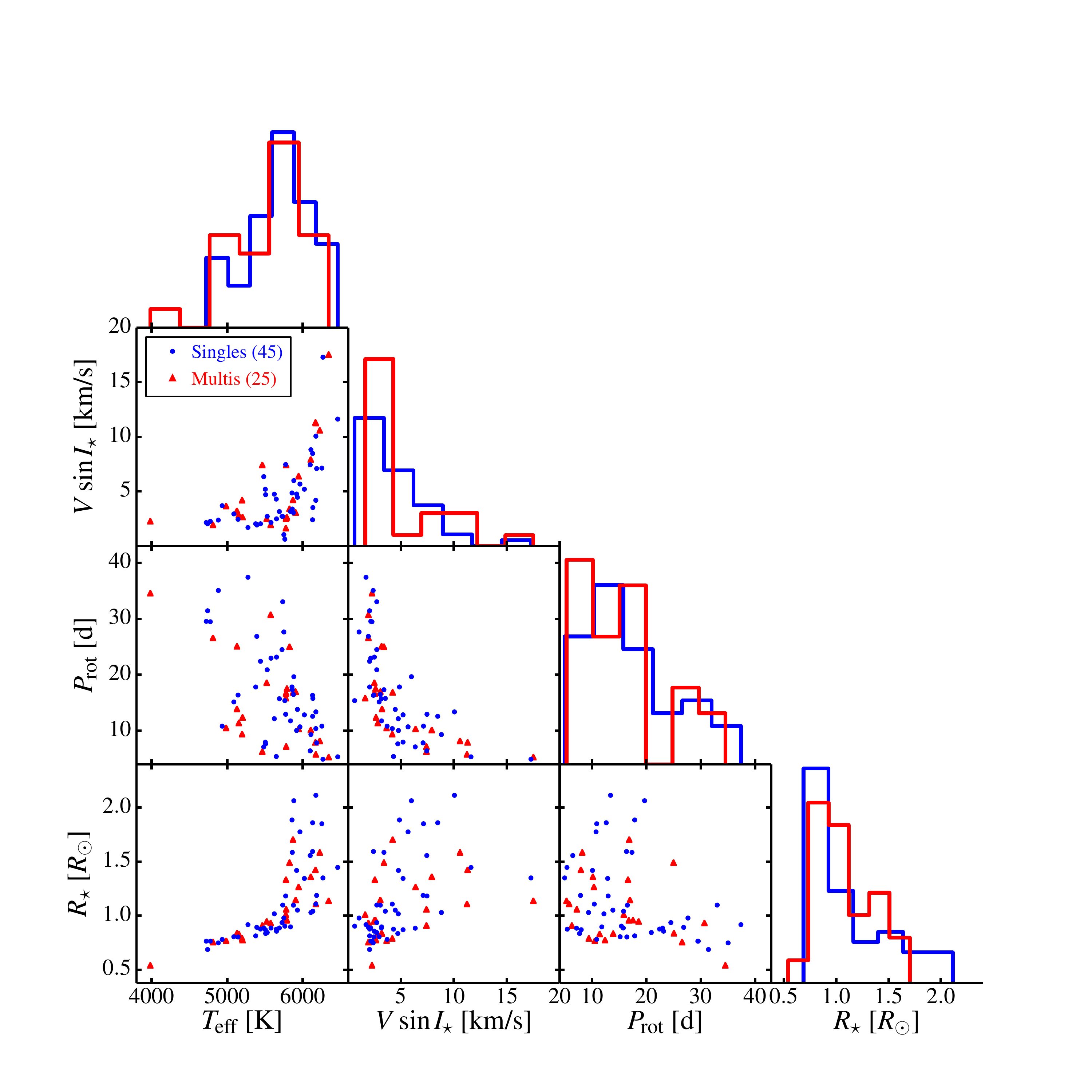} 
   \caption{
Properties of the sample of KOIs included in this study. The singles and multis have host
stars with similar properties.
The smallest planets ($R < 1.5 R_\oplus$) are almost all in multiplanet systems, whereas the largest planets ($R > 10 R_\oplus$) are all in single-transiting systems.
}
      \label{fig:propdiag}
\end{figure*}

\section{Application to real data}
\label{sec:real-data}

\subsection{Sample selection}

\citet{hirano2012b} presented measurements of $\prot$, $R_\star$, and $\vsini$ for 10 \Kepler\ objects of interest (KOIs). Three of the 10 were problematic cases: two of them (KOI-42 and KOI-279) have stellar companions which may have contaminated the spectra, and another one (KOI-1463) is likely a stellar eclipsing binary. This left 7 reliable inclination determinations which we include in our sample.

\citet{hirano2014} presented similar results for an additional 25 KOIs. Most of these KOIs are still ``candidates'' rather than confirmed planets. For each of these candidates we calculated the probability that it is an astrophysical false positive using the method of \citet{morton2012}.  We found that KOI-1615, one of the single-transiting systems from \citet{hirano2014}, is a likely false positive, and consequently we did not include it in our sample.  In addition, we dropped KOI-1835 because the radius is poorly constrained; \citet{hirano2014} quote the radius as $0.832^{+1.216}_{-0.041}~R_\odot$. This left 23 KOIs from \citet{hirano2014} that we included in our sample.

In addition, we found an additional 41 KOIs for which both the spectroscopic parameters and the rotation period have been reported in the literature \citep{buchhave2012,walkowicz2013,mcquillan2014}, and which are not likely to be false positives.  One special case was KOI-975, for which \cite{buchhave2012} reported $\vsini$ of 11.3~km~s$^{-1}$, while the \Kepler\ Community Follow-up Observing Program\footnote{\url{https://cfop.ipac.caltech.edu}} reports several spectroscopic measurements of this star with a much lower value of 7.5~km~s$^{-1}$.  Since the estimated $V$ for this star is also about 7.5~km~s$^{-1}$, we assumed that the \cite{buchhave2012} result was in error and the actual $\vsini$ for this star is $7.5\pm 1.0$~km~s$^{-1}$.  Another system, KOI-244, has a measured $\vsini$ of 9.5~km~s$^{-1}$ \citep{albrecht2013} and $2\pi R_\star/P_{\rm rot}= 2.9$~km~s$^{-1}$ which is physically impossible. We suspect the rotation period measurement is in error, perhaps due to a blended stellar companion; we omitted this star from our sample. This left a total of \Ntot\ KOIs in our sample, for which the salient properties are given in Table \ref{table:sample}.

Of these stars, \Nsingle\ host only a single transiting planet, and \Nmulti\ host multiple transiting planets. This includes several cases that were treated as single-transiting systems by \citet{hirano2014}, but for which additional transiting planets have since been detected.  For the stellar radii, we adopt the values reported by \citet{hirano2012b} or \citet{hirano2014} for the stars taken from those studies, and we adopt the values reported by \citet{huber2014} for the remainder. The properties of this sample are illustrated in Figure~\ref{fig:propdiag}. Importantly for our comparative study, the host star properties of the single and multiple systems are very similar, apart from the number of transit candidates; they span the same range of effective temperature, $\vsini$, and $P_{\rm rot}$.  The planet properties, however, do show differences apart from multiplicity: planets in multiple systems tend to be systematically smaller than planets in single systems, a trend which has been noted by \cite{lissauer2011,latham2011} and \cite{johansen2012}, among others.  We proceed to investigate the probability distribution of $\k$ for the single-KOI and multiple-KOI stars, and to see if there is a discernible difference in the stellar obliquity distributions of these two populations.

\subsection{Results}
\label{sec:results}

The first step was to calculate the $\cos I_\star$ distribution for each star, which requires probability distributions for $\veq$ and $\vsini$ (Equation \ref{eq:cosilhood}). While $\veq$ may be simply determined as $2\pi R_\star/P_{\rm rot}$ as we have said earlier, small corrections are needed due to differential rotation. If a star is differentially rotating and the spots are not on the equator, then the photometric variations will represent the rotation period at some nonzero latitude, which is probably longer than the equatorial rotation period (as is the case for the Sun).

\citet{hirano2012b} and \citet{hirano2014} addressed this complication this by adding an additional term into the error budget for $\veq$, assuming the differential rotation prescription as a function of latitude $\ell$,
\begin{equation}
\label{eq:diffrot}
P_{\rm rot}(\ell) = \frac{P_{\rm rot,eq}}{1 - \alpha \sin^2 \ell},
\end{equation}
taken from \citet{colliercameron2007}. They further assumed that the spot latitudes are $\ell = 20^\circ \pm 20^\circ$ and that the strength of differential rotation is $\alpha = 0.23$, as in the Sun. We adopted these same assumptions regarding differential rotation, but rather than trying to quantify the uncertainty by adding a systematic error term, we instead used a Monte Carlo simulation for each star: we populated the star with spots according to the assumed latitude distribution, drew rotation periods and stellar radii according to the measurements and their uncertainties (i.e., taking into account the actual posterior for the stellar radius), and calculated the equatorial rotational velocity implied by each trial, under the given assumptions regarding differential rotation. We used the resulting distribution for $\veq$ as the posterior for $\veq$ in Equation~(\ref{eq:cosilhood}).

A similar complication arises in the measurement of $\vsini$ based on spectral line profiles. Using mock data, \citet{hirano2014} demonstrated that models of spectral line profiles that ignore differential rotation will typically result in $\vsini$ values that underestimate the true values by a factor of $\approx 1 - \alpha/2$. We therefore corrected the $\vsini$ values of \citet{buchhave2012} by dividing by this factor, to accord with the already-corrected values we took from \citet{hirano2012b} and \citet{hirano2014}. We adopted the same uncertainties in $\vsini$ that were reported in the literature.

\begin{figure}[b!]
   \centering
   \includegraphics[width=3.5in]{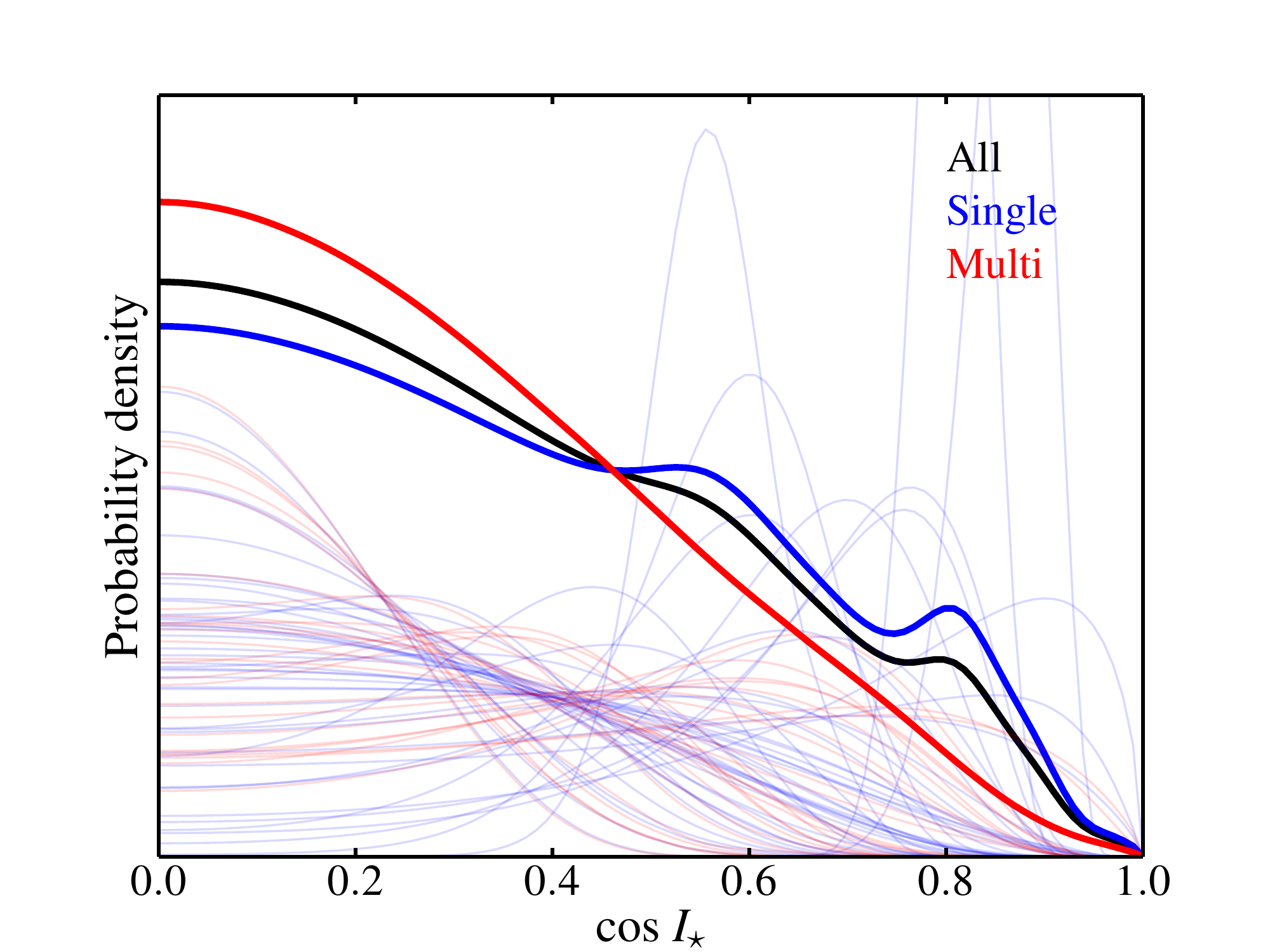} 
   \caption{The posterior probability distributions for $\cos I_\star$ for all 70
stars considered in this study. Visually, there appears to be a slight preference for the multiple-KOI systems to have lower $\cos I_\star$ than the single-KOI systems,
but it is difficult to quantify this effect by simple inspection
of the average posteriors.}
      \label{fig:avgposteriors}
\end{figure}

With probability distributions for $\veq$ and $\vsini$ in hand, we calculated the $\cos I_\star$ posterior for each KOI. Figure~\ref{fig:avgposteriors} shows these individual posteriors as well as the overall average posterior and the average posteriors for the single-KOI and multiple-KOI systems.  The average posteriors are shown here for comparision to Figure~9 of \citet{hirano2014}; we did not use the average posteriors directly for inference.
The last column in Table \ref{table:sample} gives the 95\%-confidence upper limit on $I_\star$,
based on the posterior.
The dozen systems that have the most constraining upper limits are highlighted in bold,
for the convenience of observers who may want to follow up with additional observations.
These include 11 single-transiting systems and one multi-transiting sytsem (KOI~2261).

\begin{figure}[t!]
   \centering
   \includegraphics[width=3.5in]{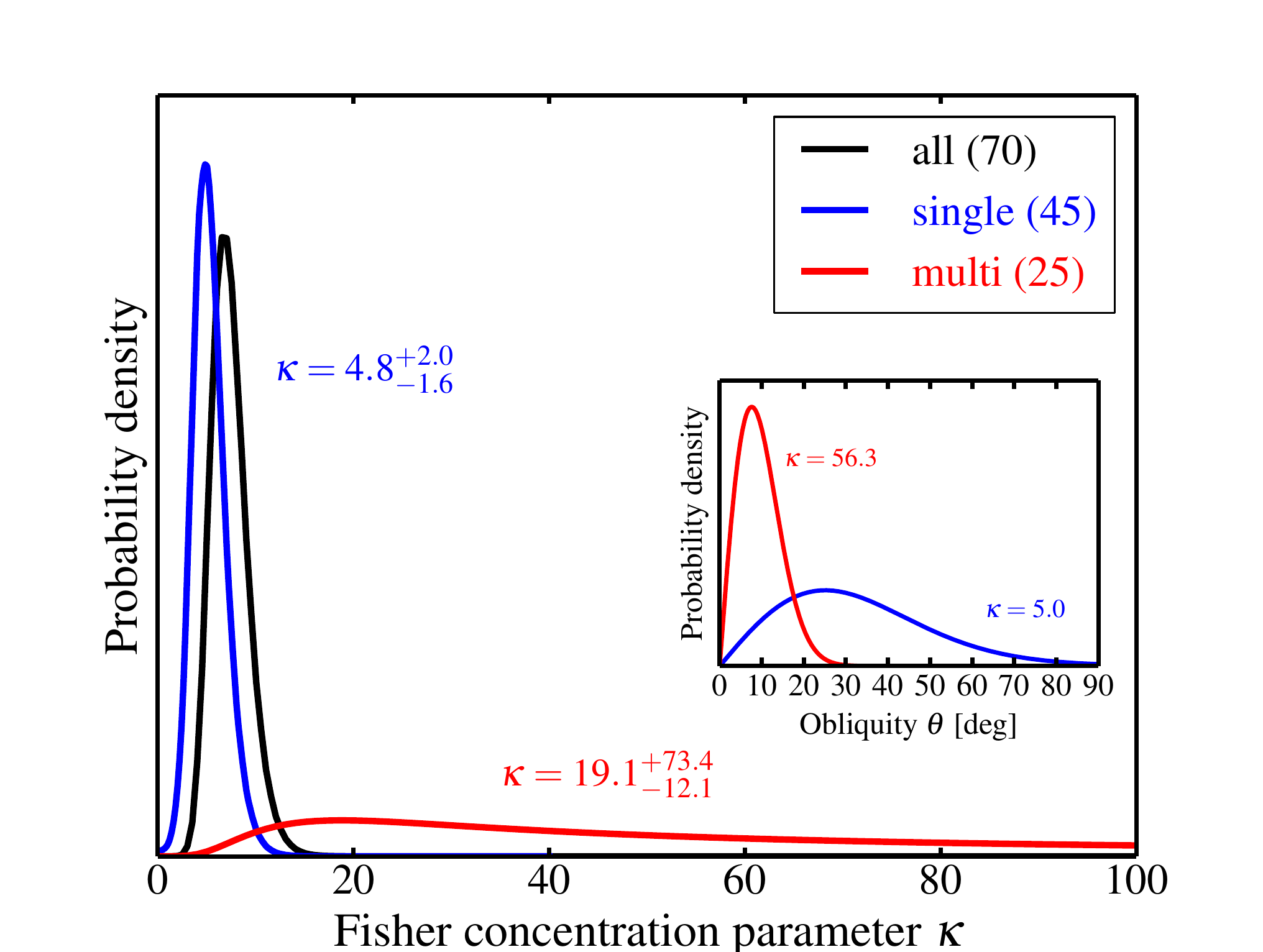} 
   \caption{The posterior probability distribution of the Fisher concentration
parameter $\k$, conditioned on the measurements of $\vsini$, $R_\star$, and $P_{\rm rot}$.
The black curve uses data from all \Ntot\ KOIs considered in this study.
The red curve is the subset that host multiple transiting candidates.
The blue curve is the subset with only one detected transiting candidate.
The multis have a significantly higher $\kappa$, corresponding
to lower obliquities (a distribution more concentrated around zero).
The inset plot shows the obliquity distributions for the singles and multis corresponding to the median $\k$ values from each posterior.}
      \label{fig:kappapost}
\end{figure}

We then proceeded to infer the posterior probability distribution for $\k$, using 1000 samples\footnote{\cite{hogg2010} demonstrated that $K\sim 50$ samples was sufficient for reliably inferring the exoplanet eccentricity distribution, and \cite{dfm2014} used $K=256$ samples in their application of this prescription.  We confirm that repeating our analysis multiple times using $K=1000$ yields negligibly different results due to sampling variance.} generated from each $\cos I$ distribution.  Figure~\ref{fig:kappapost} shows the results for the entire sample (black), and for the single-transit (blue) and multiple-transit (red) subsamples. There is a significant difference between the two subsamples: the stars with multiple transiting candidates have a higher $\kappa$ (maximum-posterior $\k=19.1$) and are therefore more concentrated toward low obliquities. The stars with only one transiting candidate show a broader obliquity distribution (maximum-posterior $\k=4.8$).   

\subsection{Significance test}

While the above analysis indicates that the multi-transit systems in our sample have lower obliquities than the single-transit systems, we would like to understand the statistical significance of the differnece.  How likely is this difference to have originated simply by chance, rather than reflecting intrinsic differences between the two populations?

To address this question, we repeated the analysis of \S\ref{sec:results} 1000 times. In each iteration we randomly assigned \Nsingle\ of the \Ntot\ stars in our sample to ``group $A$'' and \Nmulti\ to ``group $B$'' (to match the sample sizes of the single- and multi-transiting systems) and calculated the $\k$ posterior distributions for each group.  We then calculated how often these $A$ and $B$ $\k$ posteriors are ``as different'' as the actual single- and multi-transiting samples.

\begin{figure}[t!]
   \centering
   \includegraphics[width=3.5in]{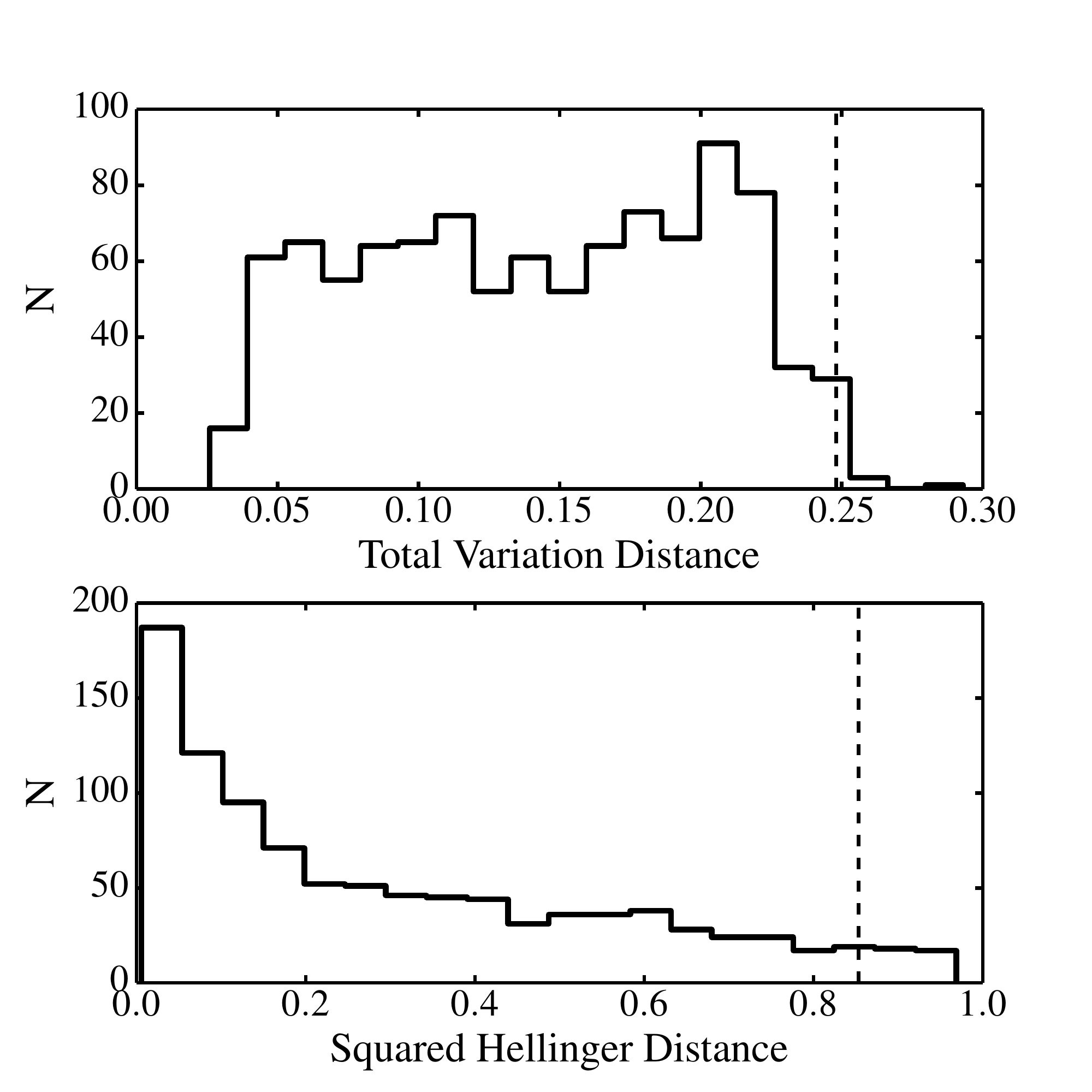} 
   \caption{Probability density function distance metrics evaluated on the $\k$ distributions calculated from two subsets of the stars in this study, for both 1000 randomized trials (histogram) and the single-/multi-transiting subsets from Figure \ref{fig:kappapost} (vertical dashed line).  The more different the two $\k$ probability distributions, the larger the value of the distance metric.  There is a $\sim$1-3\% chance that splitting the \Ntot\ stars in our study into two random subsets of \Nsingle\ and \Nmulti\ could result in $\k$ distributions as different as we measure for the single-/multi-transiting split.  If there is indeed a true distinction between the obliquity distributions of single- and multi-transiting KOIs, we expect this probability to decrease to $<$1\% with a larger sample size.}
      \label{fig:distances}
\end{figure}

For the precise meaning of ``as different'', we used two different metrics of the distance between probability density functions, calculating each metric both for the single/multi subsets shown in Figure \ref{fig:kappapost} and for each of the $A$/$B$ subsets. The metrics we used are the total variation distance\footnote{\url{http://en.wikipedia.org/wiki/Total_variation_distance_of_probability_measures}} and the squared Hellinger distance\footnote{\url{http://en.wikipedia.org/wiki/Hellinger_distance}}, defined as follows:
\begin{equation}
\label{eq:tvd}
\delta_{\rm TVD}(p_A,p_B) = \max \left(\left| p_A(\k) - p_B(\k)\right|\right),
\end{equation}
and
\begin{equation}
\label{eq:hellinger}
\delta_{\rm H^2}(p_A,p_B) = 1 - \int \sqrt{p_A(\k) p_B(\k)} d\k.
\end{equation}
Of the 1000 randomized $A$/$B$ trials, only 15 have a larger $\delta_{\rm TVD}$ than the single/multi split, and 36 have a larger $\delta_{\rm H^2}$. These results are summarized in Figure \ref{fig:distances}. The impliciation of this experiment is that if the observed obliquity variation between the two sets of
systems were simply a statistical fluke and had nothing to do with being a single- or multi-transiting system,
the chance of observing the two sets to be as different as actually observed
is approximately 1.5\% or 3.6\%, depending on the chosen metric for differences between
probability distributions.
While this null-hypothesis probability is not completely negligible,
it does suggest a true distinction between the two populations that
should be studied further with a larger sample size.

\subsection{Multiplicity, or Planet Radius?}

We chose to divide the systems into singles and multis, but as noted previously, the observed multiplicity is also correlated with planet radius: multiple-transiting systems tend to harbor smaller planets.  This makes it difficult to ascertain whether the key difference between the two samples is multiplicity, or planet radius.  For this reason we explored whether we could control for planet radius while also assessing differences in $\k$.

To this end, we repeated the analysis of \S\ref{sec:results} for the subset of stars for which the minimum planet radius obeys $1.3 < R_p < 10 R_\odot$.  The results of analyzing this subset (39 single-transiting and 15 multi-transiting systems) are qualitatively similar to the full-sample results: the maximum-posterior values of $\k$ for the single- and multi-transiting systems are 4.2 and 9.1, respectively. As before, the multis appear to be more well-aligned, even after controlling for planet radius.  However, the difference between the two distributions for these subsets is less significant than in the full sample. The null-hypothesis probability is 3.3\% using $\delta_{\rm TVD}$ as the metric for differences between probability distributions, and 16\% using $\delta_{\rm H^2}$.  For this reason we cannot draw any significant conclusions from this investigation.  Instead, we simply acknowledge that disentangling the effects of planet size and multiplicity will be an important goal of future studies with a larger sample size.

\section{Discussion and Conclusions}

We have presented a framework for characterizing the obliquity distribution of a population of stars with transiting planets, based on measurements of $\vsini$, $R_\star$, and $P_{\rm rot}$. The obliquities are assumed to obey a Fisher distribution, and hierarchical Bayesian inference is used to derive the posterior for the Fisher concentration parameter $\kappa$.  Application of this framework to \Ntot\ \Kepler\ systems implies that systems in which \Kepler\ sees multiple transiting planets tend to have lower obliquities than systems in which \Kepler\ sees only a single planet.

This is one of only few constraints that have been obtained for the obliquities of planet-hosting stars, apart from stars with hot Jupiters. It also adds to other suggestions that the \Kepler\ singles and \Kepler\ multis cannot be simultaneously explained by a single underlying population. For example, \cite{lissauer2011} and \cite{johansen2012} concluded that there is an excess of single-candidate systems over what would be expected from simple extrapolation from the numbers of multiple systems.  In particular, \citet{johansen2012} show that while the relative numbers of double- and triple-candidate systems can be well explained by typical mutual inclinations of $\sim$5$^\circ$, this same distribution can explain only about one-third of the single systems.  This suggests that single-transiting systems should be considered as a separate population from the multiples, a proposition that they called the ``\Kepler\ dichotomy.''

\citet{johansen2012} also proposed an explanation for this dichotomy: many of the single-transiting systems harbor a large planet in the sub-AU regime which inhibited the formation of additional close-in planets. This suggestion was also put forward by \citet{latham2011}. However, an alternative scenario is that single-transiting systems are a separate population of compact multiple-planet systems that have significantly larger mutual inclinations. Such ``dynamically hot'' systems would also be likely to have high stellar obliquities, and in that sense our finding that the singles have a broader obliquity distribution is compatible with this alternative scenario. Thus, we regard the results as tentative evidence for a population of more highly mutually inclined planetary systems, distinct from the ``pancake-flat'' compact multis that have received wide attention \citep{fang2012}.

Though our findings are suggestive, the statistical significance is relatively modest, and there remains a $\approx$1-4\% chance that the results we observe could have occurred by chance, even if obliquity variations were not related to whether systems were single- or multi-transiting.  We also note that our conclusions are predicated on parametrizing the obliquity distributions as Fisher, or spherical-normal, distributions.  This seems like a natural starting point for this type of investigation but there is no physical reason why the Fisher model must be correct.  If, for example, the obliquity distribution of single-transiting systems were bi-modal (or a superposition of two Fisher-like distributions), then a single-Fisher model would poorly characterize the situation.  It would therefore be desirable to analyze the obliquity distributions non-parametrically, without the need to assume a functional form.  However, due to the nature of this analysis, where the only available data delivers broad and typically asymmetric posteriors for each system, two-sample tests such as the Kolmogorov-Smirnov or Anderson-Darling tests---which are often used to quantify whether two observed data sets originate from the same or different underlying distributions---are not applicable.  Extension of the concept of the two-sample test to this particular scenario would be a desirable goal, but is beyond the scope of this work.

Furthermore, given the present small sample, we are unable to divide the sample further to see if the difference between the two subsamples is related to any particular stellar or planetary property.  We note in particular that it would be desirable to disentagle the effects of planet radius from whether a system is single- or multi-transiting, given the correlation between these two properties.  Projects are underway to analyze high-resolution spectra for much larger samples of KOIs (E.\ Petigura, priv.\ comm.\, 2014). These large samples will greatly increase the power of this technique and enable further investigation into obliquity differences between different subsamples of planetary systems.

\acknowledgments We are very grateful to Dan Fabrycky, Eric Feigelsen, Dan Foreman-Mackey, David Hogg, John Johnson,
Robert Lupton,
Erik Petigura, Roberto Sanchis-Ojeda, Kevin Schlaufman, Scott Tremaine, Ed Turner, and the anonymous referee for helpful discussions and suggestions. This work was supported by the \Kepler\ Participating Scientist Program (NNX14AE11G, NNX12AC76G), the NASA Origins program (NNX11AG85G) and the NSF (Grant No.\ 1108595).

\bibliographystyle{apj}
\bibliography{allrefs}

\tabletypesize{\scriptsize}
\begin{deluxetable*}{c|cccccc}
\tablecaption{Key properties of the sample of KOIs investigated in this work.}
\tablewidth{0pt}
\tablehead{
\colhead{KOI star} & \colhead{$R_\star$ [$R_\odot$]} & \colhead{$P_{\rm rot}$ [d]} & \colhead{\tablenotemark{$\dagger$}$\vsini$ [km s$^{-1}$]} & \colhead{$N_{\rm cands}$}  & \colhead{$I_\star$, 95\% UL [deg]}
}
\startdata

K00003 & \tablenotemark{a}$0.77 \pm^{0.03}_{0.02}$ & \tablenotemark{b}$29.47 \pm 0.13$ & \tablenotemark{a}$2.26 \pm 0.50$ & 1 & 89.4\\
K00041 & \tablenotemark{a}$1.49 \pm^{0.04}_{0.04}$ & \tablenotemark{b}$24.99 \pm 2.19$ & \tablenotemark{a}$3.39 \pm 0.50$ & 3 & 89.4\\
\bf{K00063} & \tablenotemark{a}$\boldsymbol{0.88 \pm^{0.12}_{0.03}}$ & \tablenotemark{b}$\boldsymbol{5.41 \pm 0.00}$ & \tablenotemark{a}$\boldsymbol{4.29 \pm 0.50}$ & \bf{1} & \bf{38.9}\\
K00082 & \tablenotemark{a}$0.76 \pm^{0.03}_{0.04}$ & \tablenotemark{b}$26.57 \pm 0.15$ & \tablenotemark{a}$1.92 \pm 0.50$ & 5 & 88.8\\
K00085 & \tablenotemark{a}$1.42 \pm^{0.02}_{0.02}$ & \tablenotemark{b}$7.91 \pm 0.15$ & \tablenotemark{a}$11.30 \pm 0.50$ & 3 & 90.0\\
K00103 & \tablenotemark{a}$0.86 \pm^{0.14}_{0.03}$ & \tablenotemark{b}$23.16 \pm 0.15$ & \tablenotemark{a}$2.49 \pm 0.50$ & 1 & 89.4\\
\bf{K00107} & \tablenotemark{a}$\boldsymbol{1.59 \pm^{0.06}_{0.06}}$ & \tablenotemark{b}$\boldsymbol{17.33 \pm 0.13}$ & \tablenotemark{a}$\boldsymbol{3.39 \pm 0.50}$ & \bf{1} & \bf{73.6}\\
K00135 & \tablenotemark{a}$1.34 \pm^{0.10}_{0.11}$ & \tablenotemark{b}$12.85 \pm 0.05$ & \tablenotemark{a}$5.20 \pm 0.50$ & 1 & 88.8\\
K00139 & \tablenotemark{a}$1.15 \pm^{0.25}_{0.11}$ & \tablenotemark{b}$16.96 \pm 0.10$ & \tablenotemark{a}$3.05 \pm 0.50$ & 2 & 87.7\\
K00142 & \tablenotemark{a}$0.93 \pm^{0.16}_{0.04}$ & \tablenotemark{b}$30.69 \pm 0.38$ & \tablenotemark{a}$1.92 \pm 0.50$ & 2 & 88.8\\
K00156 & \tablenotemark{a}$0.54 \pm^{0.06}_{0.06}$ & \tablenotemark{b}$34.55 \pm 0.12$ & \tablenotemark{a}$2.26 \pm 0.50$ & 3 & 90.0\\
K00180 & \tablenotemark{d}$0.89 \pm^{0.13}_{0.03}$ & \tablenotemark{d}$15.73 \pm 0.73$ & \tablenotemark{d}$3.15 \pm 0.81$ & 1 & 88.8\\
K00203 & \tablenotemark{a}$1.02 \pm^{0.06}_{0.06}$ & \tablenotemark{b}$12.16 \pm 0.03$ & \tablenotemark{a}$4.75 \pm 0.50$ & 1 & 89.4\\
K00257 & \tablenotemark{c}$1.19 \pm^{0.02}_{0.02}$ & \tablenotemark{c}$7.85 \pm 0.05$ & \tablenotemark{c}$7.09 \pm 0.49$ & 1 & 87.1\\
\bf{K00261} & \tablenotemark{c}$\boldsymbol{0.90 \pm^{0.17}_{0.04}}$ & \tablenotemark{c}$\boldsymbol{15.38 \pm 0.30}$ & \tablenotemark{c}$\boldsymbol{0.62 \pm 1.09}$ & \bf{1} & \bf{82.5}\\
K00262 & \tablenotemark{c}$1.58 \pm^{0.03}_{0.03}$ & \tablenotemark{c}$8.17 \pm 1.22$ & \tablenotemark{c}$10.58 \pm 0.22$ & 2 & 89.4\\
\bf{K00269} & \tablenotemark{c}$\boldsymbol{1.45 \pm^{0.03}_{0.03}}$ & \tablenotemark{c}$\boldsymbol{5.35 \pm 0.14}$ & \tablenotemark{c}$\boldsymbol{11.62 \pm 0.22}$ & \bf{1} & \bf{63.6}\\
K00271 & \tablenotemark{a}$1.36 \pm^{0.04}_{0.04}$ & \tablenotemark{b}$10.12 \pm 0.21$ & \tablenotemark{a}$7.91 \pm 0.50$ & 3 & 89.4\\
K00280 & \tablenotemark{c}$1.04 \pm^{0.03}_{0.03}$ & \tablenotemark{c}$15.78 \pm 2.12$ & \tablenotemark{c}$3.52 \pm 0.50$ & 1 & 88.8\\
K00283 & \tablenotemark{a}$0.96 \pm^{0.16}_{0.04}$ & \tablenotemark{b}$17.52 \pm 0.13$ & \tablenotemark{a}$2.60 \pm 0.50$ & 2 & 88.3\\
K00285 & \tablenotemark{d}$1.70 \pm^{0.05}_{0.05}$ & \tablenotemark{d}$16.83 \pm 0.59$ & \tablenotemark{d}$4.21 \pm 0.77$ & 3 & 87.1\\
K00288 & \tablenotemark{a}$2.11 \pm^{0.04}_{0.04}$ & \tablenotemark{b}$13.38 \pm 0.10$ & \tablenotemark{a}$10.06 \pm 0.50$ & 1 & 90.0\\
K00299 & \tablenotemark{a}$0.89 \pm^{0.12}_{0.03}$ & \tablenotemark{b}$22.96 \pm 0.05$ & \tablenotemark{a}$2.15 \pm 0.50$ & 1 & 88.8\\
K00304 & \tablenotemark{d}$1.01 \pm^{0.07}_{0.09}$ & \tablenotemark{d}$15.81 \pm 2.61$ & \tablenotemark{d}$1.62 \pm 1.28$ & 2 & 87.1\\
K00305 & \tablenotemark{a}$0.77 \pm^{0.03}_{0.04}$ & \tablenotemark{b}$29.55 \pm 4.76$ & \tablenotemark{a}$2.15 \pm 0.50$ & 1 & 89.4\\
K00306 & \tablenotemark{a}$0.81 \pm^{0.11}_{0.03}$ & \tablenotemark{b}$17.82 \pm 0.02$ & \tablenotemark{a}$2.03 \pm 0.50$ & 1 & 87.7\\
K00315 & \tablenotemark{a}$0.69 \pm^{0.03}_{0.04}$ & \tablenotemark{b}$31.43 \pm 3.59$ & \tablenotemark{a}$2.03 \pm 0.50$ & 1 & 89.4\\
K00318 & \tablenotemark{a}$1.35 \pm^{0.34}_{0.15}$ & \tablenotemark{b}$4.92 \pm 0.01$ & \tablenotemark{a}$17.29 \pm 0.50$ & 1 & 89.4\\
K00319 & \tablenotemark{a}$2.06 \pm^{0.08}_{0.08}$ & \tablenotemark{b}$19.65 \pm 0.15$ & \tablenotemark{a}$5.99 \pm 0.50$ & 1 & 89.4\\
\bf{K00323} & \tablenotemark{d}$\boldsymbol{0.84 \pm^{0.11}_{0.03}}$ & \tablenotemark{d}$\boldsymbol{7.67 \pm 0.14}$ & \tablenotemark{d}$\boldsymbol{4.70 \pm 0.30}$ & \bf{1} & \bf{79.5}\\
K00340 & \tablenotemark{a}$1.18 \pm^{0.28}_{0.14}$ & \tablenotemark{b}$12.94 \pm 0.02$ & \tablenotemark{a}$7.46 \pm 0.50$ & 1 & 89.4\\
K00341 & \tablenotemark{a}$0.95 \pm^{0.17}_{0.08}$ & \tablenotemark{b}$18.55 \pm 0.05$ & \tablenotemark{a}$2.49 \pm 0.50$ & 2 & 88.3\\
K00345 & \tablenotemark{a}$0.75 \pm^{0.04}_{0.04}$ & \tablenotemark{b}$35.05 \pm 0.22$ & \tablenotemark{a}$2.37 \pm 0.50$ & 1 & 89.4\\
K00346 & \tablenotemark{a}$0.81 \pm^{0.04}_{0.06}$ & \tablenotemark{b}$15.15 \pm 0.02$ & \tablenotemark{a}$2.94 \pm 0.50$ & 1 & 88.8\\
\bf{K00355} & \tablenotemark{a}$\boldsymbol{1.11 \pm^{0.18}_{0.08}}$ & \tablenotemark{b}$\boldsymbol{10.39 \pm 1.35}$ & \tablenotemark{a}$\boldsymbol{4.18 \pm 0.50}$ & \bf{1} & \bf{85.9}\\
K00361 & \tablenotemark{a}$0.94 \pm^{0.17}_{0.05}$ & \tablenotemark{b}$24.50 \pm 3.48$ & \tablenotemark{a}$2.71 \pm 0.50$ & 1 & 88.8\\
K00367 & \tablenotemark{c}$0.98 \pm^{0.18}_{0.06}$ & \tablenotemark{c}$27.65 \pm 3.56$ & \tablenotemark{c}$1.04 \pm 0.74$ & 1 & 87.7\\
\bf{K00372} & \tablenotemark{a}$\boldsymbol{0.90 \pm^{0.13}_{0.04}}$ & \tablenotemark{b}$\boldsymbol{11.77 \pm 0.02}$ & \tablenotemark{a}$\boldsymbol{3.16 \pm 0.50}$ & \bf{1} & \bf{85.9}\\
K00377 & \tablenotemark{a}$0.96 \pm^{0.18}_{0.05}$ & \tablenotemark{b}$16.75 \pm 0.08$ & \tablenotemark{a}$2.49 \pm 0.50$ & 3 & 87.7\\
K00632 & \tablenotemark{a}$0.88 \pm^{0.14}_{0.05}$ & \tablenotemark{b}$22.41 \pm 0.17$ & \tablenotemark{a}$2.03 \pm 0.50$ & 1 & 88.8\\
K00635 & \tablenotemark{d}$1.03 \pm^{0.14}_{0.05}$ & \tablenotemark{d}$9.33 \pm 0.56$ & \tablenotemark{d}$8.82 \pm 0.52$ & 1 & 90.0\\
K00640 & \tablenotemark{a}$0.92 \pm^{0.13}_{0.09}$ & \tablenotemark{b}$37.41 \pm 5.00$ & \tablenotemark{a}$1.69 \pm 0.50$ & 1 & 88.8\\
K00678 & \tablenotemark{d}$0.83 \pm^{0.03}_{0.05}$ & \tablenotemark{d}$13.87 \pm 0.06$ & \tablenotemark{d}$3.21 \pm 0.45$ & 2 & 88.8\\
K00683 & \tablenotemark{a}$1.10 \pm^{0.22}_{0.10}$ & \tablenotemark{b}$16.49 \pm 0.17$ & \tablenotemark{a}$3.05 \pm 0.50$ & 1 & 88.3\\
K00714 & \tablenotemark{a}$0.89 \pm^{0.13}_{0.06}$ & \tablenotemark{b}$26.86 \pm 0.28$ & \tablenotemark{a}$1.92 \pm 0.50$ & 1 & 88.8\\
K00718 & \tablenotemark{d}$1.33 \pm^{0.33}_{0.21}$ & \tablenotemark{d}$16.60 \pm 0.83$ & \tablenotemark{d}$2.53 \pm 1.26$ & 3 & 87.1\\
K00720 & \tablenotemark{d}$0.79 \pm^{0.04}_{0.04}$ & \tablenotemark{d}$9.38 \pm 0.03$ & \tablenotemark{d}$4.18 \pm 0.30$ & 4 & 88.8\\
K00896 & \tablenotemark{a}$0.84 \pm^{0.07}_{0.03}$ & \tablenotemark{b}$25.07 \pm 0.01$ & \tablenotemark{a}$3.16 \pm 0.50$ & 3 & 90.0\\
\bf{K00974} & \tablenotemark{c}$\boldsymbol{1.85 \pm^{0.04}_{0.04}}$ & \tablenotemark{c}$\boldsymbol{10.83 \pm 0.12}$ & \tablenotemark{c}$\boldsymbol{7.13 \pm 0.49}$ & \bf{1} & \bf{68.1}\\
K00975 & \tablenotemark{a}$1.86 \pm^{0.02}_{0.02}$ & \tablenotemark{b}$12.59 \pm 0.04$ & \tablenotemark{e}$8.47 \pm 1.00$ & 1 & 89.4\\
K00984 & \tablenotemark{a}$0.87 \pm^{0.08}_{0.03}$ & \tablenotemark{b}$7.98 \pm 0.01$ & \tablenotemark{a}$5.20 \pm 0.50$ & 1 & 88.3\\
K00987 & \tablenotemark{a}$0.84 \pm^{0.14}_{0.03}$ & \tablenotemark{b}$20.89 \pm 0.09$ & \tablenotemark{a}$2.71 \pm 0.50$ & 1 & 89.4\\
K00988 & \tablenotemark{d}$0.77 \pm^{0.07}_{0.03}$ & \tablenotemark{d}$12.36 \pm 0.06$ & \tablenotemark{d}$2.64 \pm 0.57$ & 2 & 87.1\\
K01150 & \tablenotemark{a}$1.10 \pm^{0.19}_{0.13}$ & \tablenotemark{b}$33.04 \pm 0.49$ & \tablenotemark{a}$2.71 \pm 0.50$ & 1 & 89.4\\
K01439 & \tablenotemark{a}$1.89 \pm^{0.35}_{0.47}$ & \tablenotemark{b}$17.83 \pm 0.45$ & \tablenotemark{a}$4.86 \pm 0.50$ & 1 & 88.3\\
K01445 & \tablenotemark{a}$1.14 \pm^{0.33}_{0.07}$ & \tablenotemark{b}$5.27 \pm 0.03$ & \tablenotemark{a}$17.51 \pm 0.50$ & 3 & 89.4\\
K01628 & \tablenotemark{d}$1.11 \pm^{0.16}_{0.08}$ & \tablenotemark{d}$5.76 \pm 0.38$ & \tablenotemark{d}$11.24 \pm 0.27$ & 3 & 89.4\\
K01779 & \tablenotemark{d}$1.06 \pm^{0.13}_{0.05}$ & \tablenotemark{d}$7.15 \pm 0.01$ & \tablenotemark{d}$7.41 \pm 0.24$ & 2 & 88.8\\
K01781 & \tablenotemark{d}$0.77 \pm^{0.05}_{0.03}$ & \tablenotemark{d}$10.47 \pm 0.08$ & \tablenotemark{d}$3.64 \pm 0.22$ & 3 & 88.8\\
K01797 & \tablenotemark{d}$0.78 \pm^{0.03}_{0.03}$ & \tablenotemark{d}$10.83 \pm 0.03$ & \tablenotemark{d}$3.69 \pm 0.23$ & 1 & 89.4\\
K01839 & \tablenotemark{d}$0.91 \pm^{0.05}_{0.06}$ & \tablenotemark{d}$6.25 \pm 0.03$ & \tablenotemark{d}$7.41 \pm 0.15$ & 2 & 89.4\\
\bf{K01890} & \tablenotemark{d}$\boldsymbol{1.56 \pm^{0.04}_{0.04}}$ & \tablenotemark{d}$\boldsymbol{6.42 \pm 0.04}$ & \tablenotemark{d}$\boldsymbol{7.44 \pm 0.49}$ & \bf{1} & \bf{42.5}\\
K01916 & \tablenotemark{d}$1.26 \pm^{0.11}_{0.09}$ & \tablenotemark{d}$10.32 \pm 0.10$ & \tablenotemark{d}$6.38 \pm 0.39$ & 3 & 89.4\\
K02001 & \tablenotemark{d}$0.80 \pm^{0.02}_{0.03}$ & \tablenotemark{d}$16.39 \pm 0.08$ & \tablenotemark{d}$2.44 \pm 0.66$ & 1 & 88.8\\
\bf{K02002} & \tablenotemark{d}$\boldsymbol{1.78 \pm^{0.27}_{0.26}}$ & \tablenotemark{d}$\boldsymbol{10.71 \pm 0.36}$ & \tablenotemark{d}$\boldsymbol{5.67 \pm 0.42}$ & \bf{1} & \bf{81.3}\\
\bf{K02026} & \tablenotemark{d}$\boldsymbol{1.42 \pm^{0.20}_{0.17}}$ & \tablenotemark{d}$\boldsymbol{10.05 \pm 0.56}$ & \tablenotemark{d}$\boldsymbol{4.76 \pm 0.50}$ & \bf{1} & \bf{77.8}\\
K02035 & \tablenotemark{d}$0.89 \pm^{0.05}_{0.02}$ & \tablenotemark{d}$7.13 \pm 0.10$ & \tablenotemark{d}$6.35 \pm 0.21$ & 1 & 89.4\\
K02087 & \tablenotemark{d}$1.05 \pm^{0.19}_{0.09}$ & \tablenotemark{d}$13.82 \pm 1.14$ & \tablenotemark{d}$4.46 \pm 0.73$ & 1 & 88.8\\
\bf{K02261} & \tablenotemark{d}$\boldsymbol{0.83 \pm^{0.03}_{0.04}}$ & \tablenotemark{d}$\boldsymbol{11.37 \pm 0.04}$ & \tablenotemark{d}$\boldsymbol{2.81 \pm 0.55}$ & \bf{2} & \bf{85.4}\\
K02636 & \tablenotemark{d}$1.59 \pm^{0.81}_{0.47}$ & \tablenotemark{d}$16.33 \pm 1.32$ & \tablenotemark{d}$2.40 \pm 1.26$ & 1 & 87.1

\enddata
\tablenotetext{$\dagger$}{After adjusting for differential rotation (see \S\ref{sec:results}).}
\tablenotetext{a}{\citet{buchhave2012}}
\tablenotetext{b}{\citet{mcquillan2014}}
\tablenotetext{c}{\citet{hirano2012b}}
\tablenotetext{d}{\citet{hirano2014}}
\tablenotetext{e}{\url{https://cfop.ipac.caltech.edu}}
\label{table:sample}
\end{deluxetable*}

\end{document}